\documentstyle[prl,multicol,aps,epsfig]{revtex}

\begin{document}
\twocolumn[\hsize\textwidth\columnwidth\hsize\csname@twocolumnfalse%
\endcsname

\title{Absence of Thermodynamic Phase Transition in a Model Glass
Former}
\author{Ludger Santen and Werner Krauth}
\address{CNRS-Laboratoire de Physique Statistique \\
Ecole Normale Sup{\'{e}}rieure,\\
24, rue Lhomond, 75231 Paris Cedex 05, France}
\maketitle
\begin{abstract} 
The glass transition can simply be viewed as the point at 
which the viscosity of a structurally disordered liquid reaches
$10^{13}$ Poise \cite{Angell}. This definition is operational but it sidesteps
fundamental controversies about the glass:  Is the transition a
purely dynamical phenomenon \cite{Goetze}? This would mean that
ergodicity gets broken, but that thermodynamic properties of the
liquid remain unchanged across the transition if determined as
thermodynamic equilibrium averages over the whole phase space.  The
opposite view \cite{Kauzmann,Gibbs,Adam,Nagel} claims that an underlying
thermodynamic phase transition is responsible for the dramatic
slowdown at the liquid-glass boundary. Such a phase transition
(which shows up in proper equilibrium phase space averages) 
would trigger the
dynamic standstill, and then get  masked by it.  A recent Monte
Carlo algorithm \cite{Dress} introduces non-local moves for hard-core
systems in a way which preserves micro-reversibility.  Here we use
this method to equilibrate a two-dimensional hard disc system far
within its glassy phase.  We show that indications of a thermodynamic
transition are lacking up to very high densities, and that the
glass is indistinguishable from the liquid on
purely thermodynamic grounds.
\end{abstract}
\vspace{0.5cm}
]

We considered polydisperse hard discs $i=1,\ldots,N $ with radii
$r_i$ ($r_i - r_{i-1} = \Delta/(N-1)$) in a square box of volume $V=L^2$
with periodic boundary conditions.  The polydispersity $\Delta/r_1$
was kept fixed and the system studied as a  function of the density
$\rho = \pi \sum_i r_i^2/L^2$. Note that hard-core systems are
athermal:  the phase diagram and the dynamics (up to a trivial
rescaling) are  independent of temperature. The external control
parameters are the density (for the [NV] ensemble) or the pressure
(for [NP]).

Conventional, local-move Monte Carlo simulations in the [NV] ensemble
were used to compute effective diffusion constants \cite{Hansen} at densities
for which the local algorithm was still ergodic, cf. fig. 1. We failed
to detect any finite-size effects by comparing runs with $256$ and
$1024$ discs at density $\rho = 0.764$, but we noticed 
strong dependence of the diffusivities on the disc size.
For each disc $i$ we found its diffusivity $D_i$ to agree very 
well with
\cite{Fuchs}
\begin{equation} 
D_i (\rho) \sim ( \rho-\rho^G_i)^{\alpha}.
\end{equation} 
For the $\sim 180$ largest discs we found no systematic 
dependence of $\rho_i^G  = \rho^G  \cong 0.805 \pm 0.01$ (with a best fit
$\alpha \sim 2.4$).
For the smallest discs  $i$, the extrapolated values for $\rho_i^G$
increased slowly.
Our result for $\rho^G$ agrees with what was found in a related
system (B. Doliwa, Diploma Thesis, University of Mainz (1999)).

We next performed simulations with the cluster Monte Carlo algorithm
\cite{Dress}.  There, groups of discs are swapped around a
`pivot' (cf. fig. 4) in a way which represents an alternative
Markov-chain sampling of the Boltzmann distribution. The correctness
of both implementations was validated by
comparing structural quantities (pair correlation functions) and
thermodynamic variables in the manifestly liquid regime, where the
local algorithm still converges well. We found that the cluster
Monte Carlo algorithm does not slow down as we pass $\rho_G$.
\begin{figure}[htbp]
\centerline{\epsfig{figure=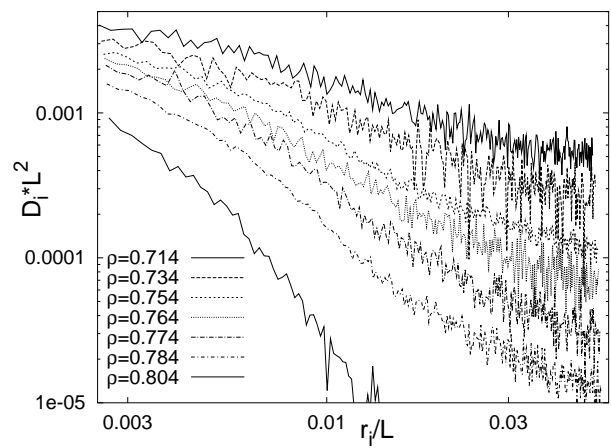,width=\columnwidth}}
\caption{\protect{Effective diffusion constants as a function of the
radius for polydisperse hard discs. 
{\bf Upper $6$ curves:} $256$ discs with polydispersity
$\Delta/r_1 = 19$ 
were simulated using the local-move Monte Carlo 
algorithm at densities at which the algorithm equilibrated. 
The effective diffusion constants were calculated {\em via} $D_i =
<({\bf x_i}(t_0+t)-{\bf x_i}(t_0))^2>/(4 t)$,  for  
$\left|{\bf x_i}(t_0+t)-{\bf x_i}(t_0) \right| \gg r_N$.
The time was measured in Monte Carlo sweeps. Only the motion relative to the
largest disc  was taken into account. These asymptotic constants
agree (up to a rescaling of the time unit) with what would be obtained
in a molecular dynamics simulation [6].
{\bf Lowest curve:} as above, but equilibrated initial configurations
were provided by the cluster Monte Carlo algorithm. We observe that
the smallest $\sim 50$ discs are able to pass through the blocked
matrix made up of the largest $\sim 200$ discs. These findings are
well consistent with the estimated values of the transition densities.}}
\label{F1}
\end{figure}

\begin{figure}[htbp]
\centerline{\epsfig{figure=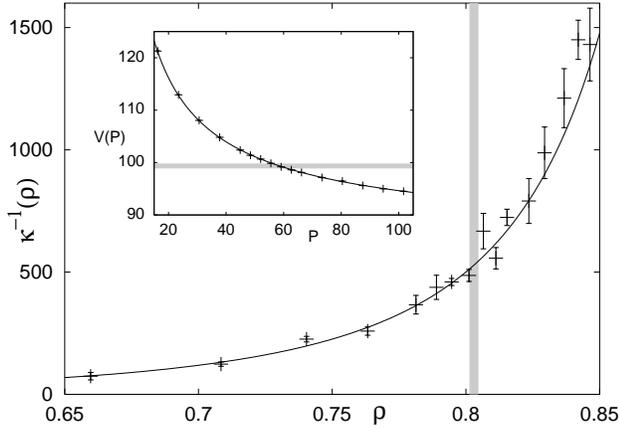,width=\columnwidth}}
\caption{\protect{Equilibrium compressibility. System as before was simulated
using the cluster Monte Carlo algorithm. Equilibrium equation of state 
(inset) and inverse compressibility $\kappa^{-1}$ (main figure)  
were obtained in the [NP] ensemble. $\kappa$ was obtained from the 
fluctuations of the volume (points with error bars) and from the 
derivative of the best $V(P)$ fit according to the functional form 
$V(P) - const  = a P^{-b}$ (Parameters of the best fit 
are $const = 84.3, a = 254, b = 0.69$). Indications of an equilibrium 
phase transition are lacking. The gray lines indicate the
localization  of the glass transition as extrapolated from fig.~1.}}
\label{F2}
\end{figure}

We used the cluster Monte Carlo algorithm to compute the  equation
of state in the [NP] ensemble \cite{Hansen}. For pressures
corresponding to the glass transition density $\rho^G$, we reached
a relative precision for $V$ of $.05\%$ during a one-week
simulation on a single processor workstation ($N=256$).
The compressibility of the system was obtained both by deriving the
algebraic fit to $V(P)$ 
(cf. inset of fig. 2) and 
by computing the volume fluctuations:
\begin{equation}
\kappa = \frac{-1}{\left< V \right>}\frac{\partial{\left< V
\right>}}{\partial{P}} = \frac{\left< V^2 \right> - \left< V
\right>^2}{\left< V \right>}   
\end{equation}

The agreement of the two methods (main fig. 2) testifies to the excellent
convergence of the cluster Monte Carlo algorithm. Some of the compressibility
data at high density were obtained from several independent runs
with identical results.  In contrast, the local Monte Carlo algorithm
yielded non-reproducible (generally lower) compressibilities.

The results presented in fig.~2 allow us to draw far-reaching
conclusions:

$i)$ In the polydisperse hard-disc system considered, the glass is
indistinguishable from the liquid on purely thermodynamic grounds,
as we find no indication of a phase transition.

$ii)$ Furthermore, the glass remains thermodynamically stable.  No
tendency towards crystallization or phase separation was detected.
At smaller values of the polydispersity $\Delta/r_1$, a crystalline
state is present, and can be readily detected in the equation of
state \cite{Alder}.  In a related two-dimensional system of  binary  mixtures,
phase separation was evidenced by the cluster Monte Carlo algorithm
\cite{Buhot}.

$iii)$  Finally, the smooth behaviour of the compressibility
as we cross $\rho^G$ 
indicates that the ergodicity of the cluster Monte Carlo algorithm 
is preserved up to the highest densities studied.

\begin{figure}[htbp]
\centerline{\epsfig{figure=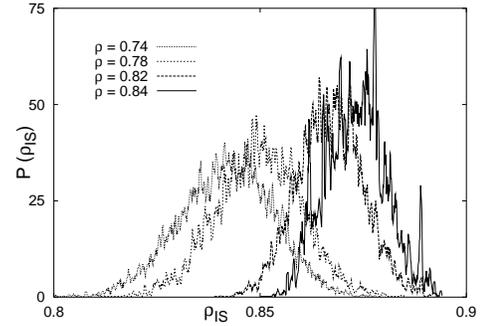,height=5cm}}
\caption{\protect{Distribution  of inherent structure densities
for $15$ discs with polydispersity $\Delta/ r_1 = 19$.  Equilibrium
configurations were generated during a long run of the cluster
Monte Carlo algorithm. The corresponding inherent structures were
obtained by iterating a rescaling of the particle radii and local
Monte Carlo moves (``rattling'') until a jamming state (of density
$\rho_{IS}$) was reached.  As $\rho$ increases, inherent structures
with higher $\rho_{IS}$ become more probable.  The same inherent structures
with very large values of $\rho_{IS}$ were found in the simulation
at $\rho=0.82$ and $\rho=0.84$.}}
\label{F3}
\end{figure}

Beyond the above point $iii)$, the cluster Monte Carlo algorithm
has passed an extremely stringent ergodicity test in a system with
$15$ polydisperse discs (where we also identified a density $\rho^G$
of complete dynamical standstill):  A large number of equilibrated
configurations were stored during single cluster-simulation runs
in the [NV] ensemble at various densities $\rho$. For each stored
configuration we then alternated very small up-scaling of each
radius $r_i \to (1+\epsilon) r_i \quad (i = 1,\dots,N)$ with a few
{\em local} Monte Carlo moves (``rattling''). The procedure was
repeated until a jamming state (with unchanged polydispersity
$\Delta/r_1$) was reached.  This state represents an inherent
structure (IS), generalised from what is done in thermal systems
by a quench to zero temperature \cite{Stillinger,Sastry}.  Histograms
of the IS densities $\rho_{IS}$ for different values of  $\rho$
show that the algorithm explores the remaining regions of phase
space as the density $\rho$ is increased, cf. fig. 3.

\begin{figure}[htbp]
\centerline{\epsfig{figure=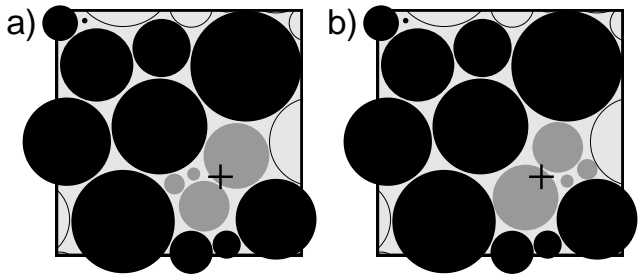,width=\columnwidth}}
\caption{\protect{Example of a cluster move at high density. The
$(+)$ denote the randomly chosen `pivot' with respect to which the
cluster (grey discs) can be flipped [7].  Periodic
boundary conditions are indicated. The inherent structure related to
the configuration a) is conjectured to yield the closest packed state
for $N=15$, $\Delta/r_1 = 19$.}}
\label{F4}
\end{figure}

A non-ergodic algorithm would get stuck in one or a few IS
(during a single run) as
regions of  phase space become {\em dynamically} inaccessible.
We observed the contrary:  almost identical  sets of IS with
extremely large values of $\rho_{IS} \gg \rho$ appeared in simulations
at different, very high, values of $\rho$ (for intermediate values
of $\rho_{IS}$, the match could no longer be achieved, simply because
of the very large number of different IS, even in a system with $15$ discs,
cf. fig. 3).  Figure 4a) shows a configuration at $\rho=
0.86$, which can be blown up into the densest IS which we were
able to find at polydispersity $\Delta/r_1 = 19$
($\rho_{IS} = 0.8938$).  This IS (as many others)
recurred frequently during a single run both at $\rho=0.84$
and at $\rho=0.82$.  We conjecture that the IS
related to fig. 4a) solves the closest-packing problem for $N=15$
and $\Delta/r_1 = 19$.

In conclusion, our results show that the new Monte Carlo
algorithm (which can be 
generalised to incorporate a smooth potential
\cite{Malherbe}) works extremely well at densities at which the
local method is totally stuck.  Besides the calculation of
thermodynamic quantities, the algorithm can also be used to study
the finite-time dynamics (starting from equilibrated samples) at
$\rho \stackrel{>}{\sim} \rho^G$.  This is how the lowest curve in
fig.~1 was obtained. As the initial condition is thermalised, we
gain access to {\em unbiased} dynamical information in a regime
which used to be out of reach of rigorous simulations because not
all discs can be properly equilibrated with the local algorithm.
Clearly, the method represents a very powerful tool to study glasses.


{\bf  Acknowledgements: \\}
We thank  C. Alba-Simionesco, J.-L. Barrat and A. Heuer for very
useful discussions. L.~S. acknowledges support from the Deutsche
Forschungsgemeinschaft under Grant No. SA864/1-1.\\ 

Correspondence and request for materials should be addressed to 
W.~K. (krauth@physique.ens.fr)

\end{document}